\documentclass[prb,superscriptaddress,amsfonts,twocolumn,showpacs,floatfix]{revtex4}

\usepackage{amsmath}
\usepackage{amsfonts}
\usepackage{amssymb}
\usepackage{bm}
\usepackage{tabularx}
\usepackage{graphicx,color}

\def\Hc2{H_\mathrm{c2}}
\def\Tc{T_\mathrm{c}}

\def\Pc{P_{\parallel c}}
\def\Pa{P_{\parallel a}}
\newcommand{\sub}[1]{\ensuremath{_{\mbox{\protect\scalebox{0.7}{#1}}}}}
\newcommand{\sro}{Sr\sub{2}{RuO\sub{4}}}
\newcommand{\SRO}{Sr\sub{3}{Ru\sub{2}O\sub{7}}}
\newcommand{\chiac}{\chi\sub{ac}}
\newcommand{\chidc}{\chi\sub{dc}}
\newcommand{\hac}{\ensuremath{H\sub{ac}}}
\newcommand{\hdc}{\ensuremath{H\sub{dc}}}

\usepackage{txfonts}

\begin{document}

\title{Higher-$\Tc$ superconducting phase in Sr$_2$RuO$_4$ induced by uniaxial pressure}

\author{Shunichiro Kittaka}
\altaffiliation[Present address: ]
{Institute for Solid State Physics, University of Tokyo, Kashiwa 277-8581, Japan}
\affiliation{Department of Physics, Graduate School of Science, Kyoto University, Kyoto 606-8502, Japan}
\author{Haruka Taniguchi}
\affiliation{Department of Physics, Graduate School of Science, Kyoto University, Kyoto 606-8502, Japan}
\author{Shingo Yonezawa}
\affiliation{Department of Physics, Graduate School of Science, Kyoto University, Kyoto 606-8502, Japan}
\author{Hiroshi Yaguchi}
\affiliation{Department of Physics, Graduate School of Science, Kyoto University, Kyoto 606-8502, Japan}
\affiliation{Department of Physics, Faculty of Science and Technology, Tokyo University of Science, Chiba 278-8510, Japan}
\author{Yoshiteru Maeno}
\affiliation{Department of Physics, Graduate School of Science, Kyoto University, Kyoto 606-8502, Japan}

\date{\today}

\begin{abstract}
We have investigated uniaxial pressure effect on superconductivity of \textit{pure} \sro, 
whose intrinsic superconducting transition temperature $\Tc$ is 1.5~K.
It was revealed that a very low uniaxial pressure along the $c$ axis, only 0.2~GPa, 
induces superconductivity with the onset $\Tc$ above 3~K.
The present results indicate that \textit{pure} \sro\ has two superconducting phases 
with $\Tc=1.5$~K and with varying $\Tc$ up to 3.2~K. 
The latter phase exhibits unusual features and 
is attributable to anisotropic crystal distortions beyond the elastic limit. 
\end{abstract}


\maketitle

The application of uniaxial pressure (UP) can be a powerful tool to control superconductivity (SC) as well as the electronic structure
through anisotropic distortions in the crystal lattice~\cite{Dix2009PRL,Campos1995PRB,Maesato2001PRB,Jin1992PRL}.
The layered perovskite ruthenate \sro, 
for which convincing evidence has been accumulated in favor of spin-triplet SC~\cite{Maeno1994Nature,Mackenzie2003RMP},
has an undistorted tetragonal structure.
It has been experimentally and theoretically revealed that 
the electronic states of \sro\ and its related materials are expected to drastically change with 
an anisotropic distortion~\cite{Braden1998PRB,Matzdorf2000Science,Fang2001PRB,Friedt2001PRB,Nakatsuji2000PRL,Nakamura2002PRB,Ikeda2004JPSJ}.
Thus, \sro\ is one of the materials expected to have a prominent UP effect on its SC. 

The intrinsic superconducting transition temperature $\Tc$ of \sro\ was revealed to be 1.5~K for crystals with the best quality~\cite{Mackenzie2003RMP}.
Hydrostatic pressure~\cite{Shirakawa1997PRB,Forsythe2002PRL}, as well as a small amount of impurities or defects~\cite{Mackenzie1998PRL,Mao1999PRB}, 
is known to suppress $\Tc$. 
In contrast, the enhancement of $\Tc$ was reported in the \sro-Ru eutectic system (the onset $\Tc \sim 3-3.5$~K)~\cite{Maeno1998PRL,Ando1999JPSJ,Kittaka2009JPSJ} and
a submicron \sro\ single crystal (the onset $\Tc \sim 1.8$~K)~\cite{Nobukane2009SSC}.
At present, the mechanisms of the enhancement of $\Tc$ remain unresolved.
In the eutectic system, lamellae of Ru metal ($\Tc=0.49$~K) with approximate dimensions of $10 \times 10 \times 1$~$\muup$m$^3$ 
are embedded with a stripe pattern.
It has been established that the 3-K SC with a tiny volume fraction occurs in the \sro\ region
in the vicinity of the \sro-Ru interface~\cite{Yaguchi2003PRB,Kawamura2005JPSJ}. 
One possible scenario is that 
anisotropic distortions in \sro, e.g. induced by the presence of Ru, enhance its $\Tc$ significantly~\cite{Ando1999JPSJ}.

Although UP experiments on \sro\ have not been reported, 
the change in $\Tc$ with UP along the interlayer $c$ and in-plane $a$ axes, $\Pc$ and $\Pa$, in the elastic limit 
was predicted from the ultrasonic experiments combined with the Ehrenfest relations~\cite{Okuda2002JPSJ}:
$(1/\Tc) (\mathrm{d}\Tc/\mathrm{d}\Pc)=(0.7 \pm 0.2)$~GPa$^{-1}$ and
$(1/\Tc) (\mathrm{d}\Tc/\mathrm{d}\Pa)=-(0.85 \pm 0.05)$~GPa$^{-1}$.
Qualitatively the same UP effects were theoretically predicted 
based on the change in the density of states at the Fermi level in the band 
mainly responsible for the SC~\cite{Nomura2002JPSJ-2}. 
On the basis of these predictions, $\Tc$ is expected to increase under $\Pc$.

Recently, the UP effects on interfacial 3-K SC in the eutectic system have been investigated~\cite{Kittaka2009JPSJ-2,Yaguchi2009JPCS}.
It was revealed that 
the volume fraction of the 3-K SC drastically increases by the application of UP in \textit{any} directions
while its onset $\Tc$ is almost invariant.
These findings urge the UP effect on \textit{pure} \sro\ to be investigated as well.

In this letter, we report the effect of $\Pc$ on pure \sro.
We revealed that the onset $\Tc$ of pure \sro\ is immediately enhanced to 3.2~K by $\Pc$ of only 0.2~GPa.
By comparing the effects of $\Pc$ between \textit{pure} \sro\ and \sro-Ru eutectic crystals,
it was newly revealed that 
the higher-$\Tc$ SC in the pure \sro\ sample was not induced by the development of interfacial 3-K SC 
originating from a tiny amount of Ru inclusions, 
but is indeed an intrinsic property of pure \sro.

Single crystals used in this study were grown by a floating zone method with Ru self-flux~\cite{Mao2000MRB}.
Because RuO$_2$ evaporates from the surface of melt during the growth, 
excess Ru tends to be left in the center of the melt.
Therefore, most of single crystalline rods with the best $\Tc$ (no defect at the Ru site)
contain \sro-Ru eutectic solidification in its core region and 
\textit{pure} \sro\ only around the thin surface part~\cite{Ando1999JPSJ}.
This feature, as well as the tendency to cleave easily, makes it difficult to obtain a large crystal which only contains \sro.
In particular, it is extremely difficult to prepare a pure sample suitable for the application of $\Pa$.
For the application of $\Pc$, we succeeded in preparing one large sample of practically pure \sro\ with almost no Ru inclusions (Sample~1) 
from a crystalline rod with relatively lower $\Tc$ (small amount of defects at the Ru site).
The dimensions of Sample~1 are $1.5 \times 1.4$~mm$^2$ in the $ab$ plane and 0.22~mm along the $c$ axis.
In order to reduce the amount of lattice defects and oxygen deficiencies as well as the number of Ru inclusions, 
Sample~1 was annealed in oxygen at 1 atm and 1050~$^\circ$C for a week.
Then, Sample~1 exhibits a very sharp transition at 1.34~K, as determined from the ac susceptibility.

\begin{figure}
\begin{center}
\includegraphics[width=3in]{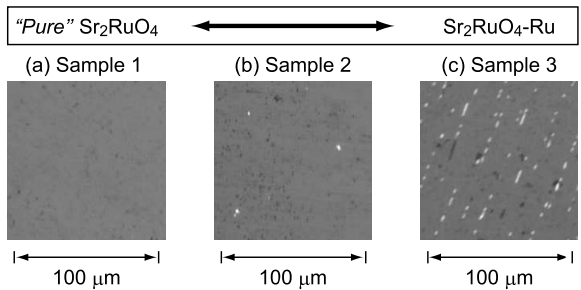}
\end{center}
\caption{
Polarized-light optical microscope images of the $ab$ planes of (a)~Samples~1, (b)~2, and (c)~3.
The dark and bright parts correspond to \sro\ and Ru, respectively.
}
\label{sample}
\end{figure}

In addition to Sample~1, 
two \sro\ samples containing a small amount of Ru inclusions and five \sro-Ru eutectic samples 
have been used to investigate the $\Pc$ effect. 
Representing those samples, magnetic susceptibility results for 
a \sro\ sample with a small amount of Ru inclusions (Sample~2), and 
a \sro-Ru eutectic sample (Sample~3) are compared with the results for Sample~1 in this letter.
The approximate dimensions of Samples~2 and 3 were 
$1.5 \times 1.5$~mm$^2$ in the $ab$ plane and 0.3~mm along the $c$ axis. 
The amount of Ru inclusions was identified from the 
polarized-light optical microscope images of the $ab$ surfaces, 
as exemplified in Figs.~\ref{sample}(a)-\ref{sample}(c).
The number density of Ru inclusions on the $ab$ surfaces for Sample~1 is less than 3~mm$^{-2}$ and about 40~mm$^{-2}$
after and before annealing, respectively, 
while the number densities are about 200~mm$^{-2}$ for Sample~2 and about 4000~mm$^{-2}$ for Sample~3.

UP was applied parallel to the shortest dimension along the $c$ axis 
using a piston-cylinder type pressure cell 
made of Cu-Be alloy with a cylindrical outer body made of hard plastic (polybenzimidazole). 
The room-temperature pressure value was confirmed to be in a reasonable agreement with low-temperature pressure value
determined by superconducting transitions of tin and lead \cite{Smith1967PR}. 
The side surfaces of the samples were covered with thin epoxy (Emerson-Cuming, Stycast 1266)
to prevent a breakdown of the sample, as described in Ref.~\cite{Kittaka2009JPSJ-2}.
The magnetization was measured down to 1.8~K with an applied dc magnetic field $\mu_0\hdc$ of 2~mT using a SQUID magnetometer (Quantum Design, model MPMS).
The background magnetization of the pressure cell was subtracted. 
The ac susceptibility $\chi_\mathrm{ac}=\chi^\prime-i \chi^{\prime \prime}$ was measured down to 0.3~K 
by a mutual-inductance technique using a lock-in amplifier (LIA) 
with a $^3$He cryostat (Oxford Instruments, model Heliox VL). 
All $\chiac$ data presented here were taken at 293~Hz.
The values of $\chi_\mathrm{ac}$ were obtained from the relation $\chi_\mathrm{ac}=iC_1V_\mathrm{LIA}/H_\mathrm{ac}+C_2$, 
where $V_\mathrm{LIA}$($=V_x+iV_y$) is the read-out voltage of LIA and $H_\mathrm{ac}$ is the magnitude of the applied ac magnetic field.
The values of $C_1$ and $C_2$ were chosen so that $\chi^\prime(\mathrm{4~K})=0$ and $\chi^\prime(\mathrm{0.3~K})=-1$ at 0~GPa for each sample; 
thus $|\chi^\prime|$ corresponds to the ac shielding fraction.
For the $\chi_\mathrm{ac}$ curves under $\Pc$, $C_1$ and $C_2$ determined at 0~GPa were used.


\begin{figure}
\begin{center}
\includegraphics[width=2.65in]{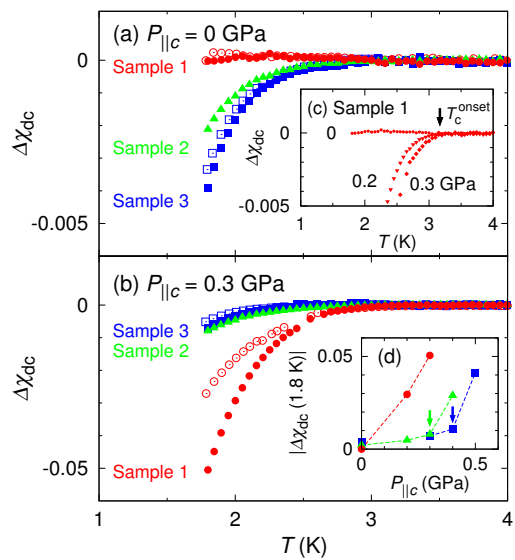}
\end{center}
\caption{
(Color Online) Temperature dependence of the dc susceptibility $\Delta\chidc$ of Samples~1 (circles), 2 (triangles), and 3 (squares) measured 
with 2~mT at (a) $\Pc=0$ and (b) 0.3~GPa. 
The applied dc field has been corrected for the remanent field in the sample space. 
Open and closed symbols indicate data taken in the FC and ZFC processes, respectively (Samples 1 and 3 only).
Note the vertical scale changes between (a) and (b).
(c)~Enlarged view near the onset for Sample~1 at different $\Pc$.
(d)~Dependence of the dc shielding fraction on $\Pc$ at 1.8~K. 
The arrows indicate critical pressure $\Pc^*$.
}
\label{MPMS}
\end{figure}

Figure~\ref{MPMS}(a) represents the temperature dependence of the superconducting dc susceptibility $\Delta\chidc=\Delta m/\mu_0\hdc$ for each sample at 0~GPa,
where $\Delta m$ is the observed magnetization change associated with the superconducting transition divided by the sample volume.
Here, the ideal value for the full Meissner state without the demagnetization correction corresponds to $\Delta\chidc=-1$;
thus $|\Delta\chidc|$ is equal to the dc shielding fraction.
At 0~GPa, there is no sign of a dc shielding signal for Sample~1 above 1.8~K,
confirming that it does not contain any eutectic part.
A weak shielding signal with the onset $\Tc$ slightly above 3~K was observed in Samples~2 and 3,
consistent with the recent study~\cite{Kittaka2009JPSJ}.
At 0~GPa, the dc shielding fraction at 1.8~K is clearly larger in Sample~3,
containing more Ru inclusions. 

Surprisingly, even at relatively low $\Pc$ of 0.2~GPa, 
SC with the onset $\Tc$ of 3.2~K, 
at which $\Delta\chidc$ deviates from zero (see Fig.~\ref{MPMS}(c)),
is induced in Sample~1.
As shown in Fig.~\ref{MPMS}(b), 
the shielding fraction in the field-cooling (FC) process is nearly half of that in the zero-field-cooling (ZFC) process at $\Pc=0.3$~GPa.
The relatively large FC shielding fraction indicates that the screening area does not contain normal-state regions largely.
Figure~\ref{MPMS}(d) demonstrates the variation of the dc shielding fraction at 1.8~K, $|\Delta\chidc(1.8~\mathrm{K})|$, under $\Pc$ for each sample.
Unexpectedly, 
as the amount of Ru inclusions is \textit{larger}, the enhancement of the shielding fraction by $\Pc$ becomes \textit{smaller};
the order of increasing shielding fractions among different samples is reversed between Figs.~\ref{MPMS}(a) and \ref{MPMS}(b).
Below 0.3~GPa, the slope of $|\Delta\chidc(1.8~\mathrm{K})|$ versus $\Pc$ for Sample~1 
is much greater than those for Samples~2 and 3.
These results strongly indicate that 
the presence of Ru inclusions does not play a positive role in the rapid enhancement of the shielding fraction under $\Pc$.
Moreover, it should be noted that the slope for Samples~2 and 3 abruptly changes at a critical pressure, 
which we designate as $\Pc^*$, of around 0.4~GPa.
Interestingly, the steep slope above $\Pc^*$ for Samples~2 and 3 is almost the same as the slope for Sample~1.

\begin{figure}
\begin{center}
\includegraphics[width=2.7in]{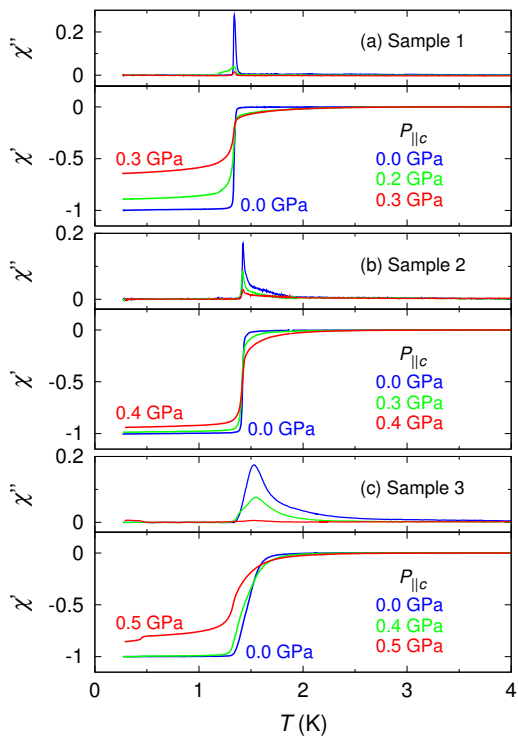}
\end{center}
\caption{
(Color Online) Temperature dependence of the real and imaginary parts of $\chiac$ under different $\Pc$
for (a) Sample~1, (b) Sample~2, and (c) Sample~3 measured with $\mu_0\hac=2$~$\muup$T-rms and 293~Hz. 
}
\label{Heliox}
\end{figure}

We have also measured the temperature dependence of $\chiac$ using the identical samples under $\Pc$
with $\mu_0H_\mathrm{ac}$ of 2~$\muup$T-rms. 
These $\chi_\mathrm{ac}$ measurements cover low temperatures down to 0.3~K and
enable us to observe the full Meissner state.
As presented in Fig.~\ref{Heliox}(a), 
Sample~1 exhibits a very sharp transition and no dissipation $\chi^{\prime\prime}$ above 1.5~K at 0~GPa. 
This sharp transition again supports that Sample~1 can be treated as a pure \sro.
By contrast, as shown in Figs.~\ref{Heliox}(b) and \ref{Heliox}(c),
broad transitions, typical of the eutectic system~\cite{Kittaka2009JPSJ}, were observed above 1.5~K for Samples~2 and 3 at 0~GPa.

By the application of $\Pc$, 
a broad signal of ac shielding with the onset $\Tc$ of up to 3.2~K grows in all samples; 
the enhancement of the ac shielding fraction 
was clearly observed above 1.5~K even for pure \sro.
These results are 
both qualitatively and quantitatively consistent with the results of the dc magnetization measurements in Fig.~\ref{MPMS}.
By contrast, at low temperatures, 
the ac shielding signal is reduced in magnitude and becomes broader with increasing $\Pc$.
It should also be noted that 
the primary 1.5-K SC part in Samples 1 and 2 is sustained with little change of its $\Tc$ under $\Pc$.
These features imply that 
a region with widely distributed $\Tc$ from zero up to above 3~K emerges, replacing some parts of the 1.5-K SC 
(i.e. two superconducting phases coexist).
Although the shielding signal near the onset $\Tc$ becomes somewhat smaller after removing uniaxial pressure 
(e.g. Fig. 4(a) of Ref.~\cite{Kittaka2009JPSJ-2}), the recovery was not complete. 
Such irreversibility suggests that the enhancement of $\Tc$ is accompanied by distortions exceeding the elastic limit.

As evident in Fig.~\ref{Heliox}(c), 
the reduction of the shielding fraction and the broadening of the transition below 1.5~K 
become particularly significant for $\Pc \ge \Pc^*$ (Fig.~\ref{MPMS}(d)).
Above $\Pc^*$, the interfacial 3-K SC around Ru inclusions also seems suppressed, because
the screening signal of the superconducting transition of Ru ($\Tc=0.49$~K)
becomes substantially strong.
Such screening signal of Ru SC was reproducibly observed in all the eutectic samples at higher $\Pc$.

\begin{figure}
\begin{center}
\includegraphics[width=2.7in]{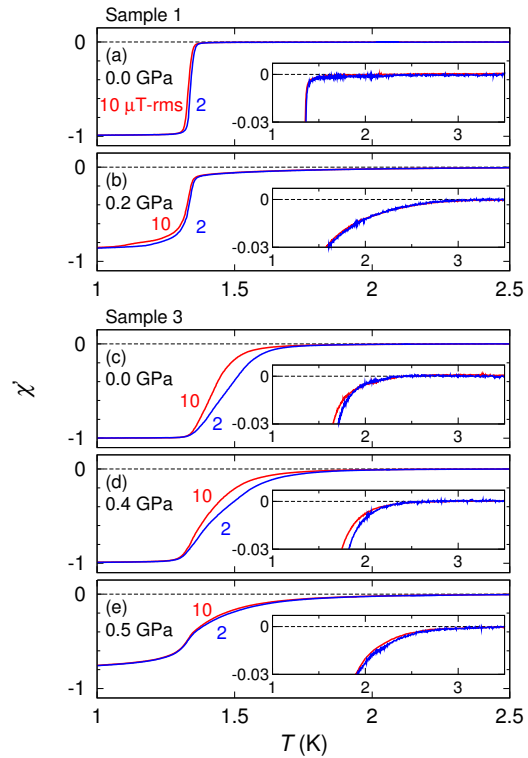}
\end{center}
\caption{
(Color Online) Temperature dependence of $\chi^\prime$ for $\mu_0H_\mathrm{ac}=2$ and 10 $\muup$T-rms at (a) $\Pc=0$ and (b) 0.2~GPa for Sample~1 and
at (c) $\Pc=0$, (d) 0.4, and (e) 0.5~GPa for Sample~3.
The insets are enlarged views near the onset.
The dashed lines mark $\chi^\prime=0$.
}
\label{hac-dep}
\end{figure}

Another important feature of the UP-induced higher-$\Tc$ SC is that the screening is insensitive to $\hac$.
As shown in Figs.~\ref{hac-dep}(a) and \ref{hac-dep}(b),
the ac shielding fraction in pure \sro\ was not affected by $\hac$ at any $\Pc$ investigated.
In contrast, the ac shielding fraction in Sample~3 was sensitively suppressed by $\hac$ of as small as 10~$\muup$T-rms
below $\Pc^* \sim 0.4$~GPa (Figs.~\ref{hac-dep}(c) and \ref{hac-dep}(d)).
This ``weak'' SC in the eutectic system is attributed to 
the formation of a Josephson network~\cite{Kittaka2009JPSJ};
interfacial 3-K SC occurring near the \sro-Ru interface penetrates into the normal-state \sro\ region
due to the proximity effect and forms Josephson-type weak links among the Ru lamellae.
In this case, because of the small critical current in the Josephson network area,
the ac field penetration is expected to become significantly deeper with increasing $\hac$.
Therefore the spatial distribution of superconducting regions is different between the two systems.

Interestingly, as shown in Fig.~\ref{hac-dep}(e), 
the $\hac$ dependence almost disappears in Sample~3 above $\Pc^*$. 
This disappearance, in conjunction with the observation in Figs.~\ref{MPMS}(d) and \ref{Heliox}(c), can be consistently understood 
if the nature of the higher-$\Tc$ SC induced by UP in pure \sro\ and 
that of SC in the eutectic crystal \textit{above} $\Pc^*$ is qualitatively the same.

From the present study,
it is most likely that UP-induced higher-$\Tc$ SC is an intrinsic property of \sro, 
because the enhancement of the higher-$\Tc$ SC by $\Pc$ is more striking in samples with smaller amount of Ru inclusions.
Let us here discuss possible mechanisms of the occurrence of the higher-$\Tc$ SC.
While hydrostatic pressure destroys SC of \sro~\cite{Shirakawa1997PRB,Forsythe2002PRL}, 
UP induces SC with $\Tc$ above 3~K.
These facts indicate that 
anisotropic distortions caused by UP in the crystal structure play essential roles in enhancing $\Tc$.
Nevertheless, the present results are essentially different from the UP effect 
predicted from the ultrasonic experiments, $\mathrm{d}\Tc/\mathrm{d}\Pc \sim 1$~K/GPa.
The observed onset $\Tc$ increases up to 3.2~K even at $\Pc$ of as small as 0.2~GPa but does not exceed 3.2~K at higher $\Pc$
(see Fig.~\ref{MPMS}(c)).

This disagreement with the hydrostatic pressure effect and the elastic-limit expectation suggests that 
a qualitative change in the electronic structure of \sro\ arises at relatively low $\Pc$ and generates SC with enhanced $\Tc$.
It is known that the electronic states of Sr$_{n+1}$Ru$_n$O$_{3n+1}$ series are significantly affected by 
the rotation, tilting, and flattening of the RuO$_6$ octahedra~\cite{Matzdorf2000Science,Friedt2001PRB,Fang2001PRB}.
For example, $\Pc$ of about 0.4~GPa changes the electronic state of the bi-layer system \SRO\ 
from a metamagnetic metal to a ferromagnetic metal~\cite{Ikeda2004JPSJ}. 
In the case of \sro, 
its crystal structure is close to an instability against the in-plane RuO$_6$ rotation,
as evidenced by the softening of the rotational phonon mode observed in inelastic neutron scattering experiments~\cite{Braden1998PRB}.
Theoretically, it was predicted that 
the RuO$_6$ rotation leads to the decrease in the band widths and 
to the increase in $\Tc$ through the enhancement of the density of states associated with the van Hove singularity of one of the bands~\cite{Sigrist2001JPSJ,Matzdorf2000Science}.
Also, the tilting and flattening of the RuO$_6$ octahedra are expected to make the band widths narrower~\cite{Fang2001PRB}.

The observed distribution of $\Tc$ in UP-induced higher-$\Tc$ SC may be explained by lattice imperfection.
If $\Tc$ depends on the magnitude of local anisotropic distortion, 
$\Tc$ would be distributed because of experimentally unavoidable inhomogeneity of the distortion.
In addition, $\Tc$ should be suppressed by the pair-breaking scatterings 
due to lattice imperfection, e.g. the dislocations or rapid spatial variation of the distortion,
leading to the reduction of the mean free path.
Note that SC is completely destroyed in pure \sro\ if the mean free path is less than about 50~nm~\cite{Mackenzie1998PRL}. 
Therefore, the higher $\Tc$ is expected only in the \sro\ region with a coherent distortion.
This scenario is consistent with the recent report on the eutectic system \cite{Ying2009PRL} 
suggesting that the enhancement of $\Tc$ occurs in the \sro\ region at least 18~nm away from the interface, 
rather than at the immediate vicinity of the interface with a lot of dislocations.

To summarize, we present the first report on the UP effect on \sro. 
We found an remarkable increase of the onset $\Tc$ from 1.5~K to 3.2~K induced by $\Pc$ below 0.2~GPa.
This strong enhancement of $\Tc$ cannot be explained by the effect of $\Pc$ in the elastic limit 
deduced from the Ehrenfest relation for \sro.
We revealed that the dc shielding fraction associated with
the UP-induced higher-$\Tc$ SC and the amount of Ru inclusions are anticorrelated.
This fact clearly indicates that the higher-$\Tc$ SC is not associated with the presence of Ru inclusions.

There remain some important issues to be clarified.
Although the spatial distribution of superconducting regions is revealed to be different 
between the UP-induced higher-$\Tc$ SC and the eutectic 3-K SC,
the values of their $\Tc$'s are surprisingly similar.
It should be clarified whether or not the origin of the enhanced $\Tc$ is similar between UP-induced SC and the eutectic SC.
Another important issue is to resolve the discrepancy 
between the results of previous hydrostatic-pressure study and of the present UP experiments.
Additional experiments under in-plane UP and the crystal-structure analysis under UP
may provide important clues,
although it is technically difficult at present to perform such experiments.

\acknowledgments
We thank F. Nakamura, N. Takeshita, T. Nakamura, H. Takatsu, M. Kriener, Y. Nakai and K. Ishida for their support and
M. Sigrist for useful discussions. 
This work is supported by a Grant-in-Aid for the Global COE program ``The Next Generation of Physics, Spun from Universality and Emergence'' 
from the MEXT of Japan and
by a Grant-in-Aid for Scientific Research from the MEXT and from the JSPS.
S. K. is supported by the JSPS.

\bibliographystyle{apsrev2}
\bibliography{C:/usr/local/share/texmf/bibref/ref091007.bib}

\begin{thebibliography}{31}
\expandafter\ifx\csname natexlab\endcsname\relax\def\natexlab#1{#1}\fi
\expandafter\ifx\csname bibnamefont\endcsname\relax
  \def\bibnamefont#1{#1}\fi
\expandafter\ifx\csname bibfnamefont\endcsname\relax
  \def\bibfnamefont#1{#1}\fi
\expandafter\ifx\csname citenamefont\endcsname\relax
  \def\citenamefont#1{#1}\fi
\expandafter\ifx\csname url\endcsname\relax
  \def\url#1{\texttt{#1}}\fi
\expandafter\ifx\csname urlprefix\endcsname\relax\def\urlprefix{URL }\fi
\providecommand{\bibinfo}[2]{#2}
\providecommand{\eprint}[2][]{\url{#2}}

\bibitem[{\citenamefont{Dix $et$~$al$.}(2009)\citenamefont{Dix, Swartz, Zieve,
  Cooley, Sayles, and Maple}}]{Dix2009PRL}
\bibinfo{author}{\bibfnamefont{O.~M.} \bibnamefont{Dix}}
  \bibnamefont{$et$~$al$.}, \bibinfo{journal}{Phys. Rev. Lett.}
  \textbf{\bibinfo{volume}{102}}, \bibinfo{pages}{197001}
  (\bibinfo{year}{2009}).

\bibitem[{\citenamefont{Campos $et$~$al$.}(1995)\citenamefont{Campos, Brooks,
  van Bentum, Perenboom, Klepper, Sandhu, Valfells, Tanaka, Kinoshita,
  Kinoshita $et$~$al$.}}]{Campos1995PRB}
\bibinfo{author}{\bibfnamefont{C.~E.} \bibnamefont{Campos}}
  \bibnamefont{$et$~$al$.}, \bibinfo{journal}{Phys. Rev. B}
  \textbf{\bibinfo{volume}{52}}, \bibinfo{pages}{R7014} (\bibinfo{year}{1995}).

\bibitem[{\citenamefont{Maesato $et$~$al$.}(2001)\citenamefont{Maesato, Kaga,
  Kondo, and Kagoshima}}]{Maesato2001PRB}
\bibinfo{author}{\bibfnamefont{M.}~\bibnamefont{Maesato}}
  \bibnamefont{$et$~$al$.}, \bibinfo{journal}{Phys. Rev. B}
  \textbf{\bibinfo{volume}{64}}, \bibinfo{pages}{155104}
  (\bibinfo{year}{2001}).

\bibitem[{\citenamefont{Jin $et$~$al$.}(1992)\citenamefont{Jin, Carter, Ellman,
  Rosenbaum, and Hinks}}]{Jin1992PRL}
\bibinfo{author}{\bibfnamefont{D.~S.} \bibnamefont{Jin}}
  \bibnamefont{$et$~$al$.}, \bibinfo{journal}{Phys. Rev. Lett.}
  \textbf{\bibinfo{volume}{68}}, \bibinfo{pages}{1597} (\bibinfo{year}{1992}).

\bibitem[{\citenamefont{Maeno $et$~$al$.}(1994)\citenamefont{Maeno, Hashimoto,
  Yoshida, Nishizaki, Fujita, Bednorz, and Lichtenberg}}]{Maeno1994Nature}
\bibinfo{author}{\bibfnamefont{Y.}~\bibnamefont{Maeno}}
  \bibnamefont{$et$~$al$.}, \bibinfo{journal}{Nature (London)}
  \textbf{\bibinfo{volume}{372}}, \bibinfo{pages}{532} (\bibinfo{year}{1994}).

\bibitem[{\citenamefont{Mackenzie and Maeno}(2003)}]{Mackenzie2003RMP}
\bibinfo{author}{\bibfnamefont{A.~P.} \bibnamefont{Mackenzie}}
  \bibnamefont{and} \bibinfo{author}{\bibfnamefont{Y.}~\bibnamefont{Maeno}},
  \bibinfo{journal}{Rev. Mod. Phys.} \textbf{\bibinfo{volume}{75}},
  \bibinfo{pages}{657} (\bibinfo{year}{2003}).

\bibitem[{\citenamefont{Braden $et$~$al$.}(1998)\citenamefont{Braden,
  Reichardt, Nishizaki, Mori, and Maeno}}]{Braden1998PRB}
\bibinfo{author}{\bibfnamefont{M.}~\bibnamefont{Braden}}
  \bibnamefont{$et$~$al$.}, \bibinfo{journal}{Phys. Rev. B}
  \textbf{\bibinfo{volume}{57}}, \bibinfo{pages}{1236} (\bibinfo{year}{1998}).

\bibitem[{\citenamefont{Matzdorf $et$~$al$.}(2000)\citenamefont{Matzdorf, Fang,
  Ismail, Zhang, Kimura, Tokura, Terakura, and Plummer}}]{Matzdorf2000Science}
\bibinfo{author}{\bibfnamefont{R.}~\bibnamefont{Matzdorf}}
  \bibnamefont{$et$~$al$.}, \bibinfo{journal}{Science}
  \textbf{\bibinfo{volume}{289}}, \bibinfo{pages}{746} (\bibinfo{year}{2000}).

\bibitem[{\citenamefont{Fang and Terakura}(2001)}]{Fang2001PRB}
\bibinfo{author}{\bibfnamefont{Z.}~\bibnamefont{Fang}} \bibnamefont{and}
  \bibinfo{author}{\bibfnamefont{K.}~\bibnamefont{Terakura}},
  \bibinfo{journal}{Phys. Rev. B} \textbf{\bibinfo{volume}{64}},
  \bibinfo{pages}{020509(R)} (\bibinfo{year}{2001}).

\bibitem[{\citenamefont{Friedt $et$~$al$.}(2001)\citenamefont{Friedt, Braden,
  Andre, Adelmann, Nakatsuji, and Maeno}}]{Friedt2001PRB}
\bibinfo{author}{\bibfnamefont{O.}~\bibnamefont{Friedt}}
  \bibnamefont{$et$~$al$.}, \bibinfo{journal}{Phys. Rev. B}
  \textbf{\bibinfo{volume}{63}}, \bibinfo{pages}{174432}
  (\bibinfo{year}{2001}).

\bibitem[{\citenamefont{Nakatsuji and Maeno}(2000)}]{Nakatsuji2000PRL}
\bibinfo{author}{\bibfnamefont{S.}~\bibnamefont{Nakatsuji}} \bibnamefont{and}
  \bibinfo{author}{\bibfnamefont{Y.}~\bibnamefont{Maeno}},
  \bibinfo{journal}{Phys. Rev. Lett.} \textbf{\bibinfo{volume}{84}},
  \bibinfo{pages}{2666} (\bibinfo{year}{2000}).

\bibitem[{\citenamefont{Nakamura $et$~$al$.}(2002)\citenamefont{Nakamura, Goko,
  Ito, Fujita, Nakatsuji, Fukazawa, Maeno, Alireza, Forsythe, and
  Julian}}]{Nakamura2002PRB}
\bibinfo{author}{\bibfnamefont{F.}~\bibnamefont{Nakamura}}
  \bibnamefont{$et$~$al$.}, \bibinfo{journal}{Phys. Rev. B}
  \textbf{\bibinfo{volume}{65}}, \bibinfo{pages}{220402(R)}
  (\bibinfo{year}{2002}).

\bibitem[{\citenamefont{Ikeda $et$~$al$.}(2004)\citenamefont{Ikeda, Shirakawa,
  Yanagisawa, Yoshida, Koikegami, Koike, Kosaka, and Uwatoko}}]{Ikeda2004JPSJ}
\bibinfo{author}{\bibfnamefont{S.-I.} \bibnamefont{Ikeda}}
  \bibnamefont{$et$~$al$.}, \bibinfo{journal}{J. Phys. Soc. Jpn.}
  \textbf{\bibinfo{volume}{73}}, \bibinfo{pages}{1322} (\bibinfo{year}{2004}).

\bibitem[{\citenamefont{Shirakawa $et$~$al$.}(1997)\citenamefont{Shirakawa,
  Murata, Nishizaki, Maeno, and Fujita}}]{Shirakawa1997PRB}
\bibinfo{author}{\bibfnamefont{N.}~\bibnamefont{Shirakawa}}
  \bibnamefont{$et$~$al$.}, \bibinfo{journal}{Phys. Rev. B}
  \textbf{\bibinfo{volume}{56}}, \bibinfo{pages}{7890} (\bibinfo{year}{1997}).

\bibitem[{\citenamefont{Forsythe $et$~$al$.}(2002)\citenamefont{Forsythe,
  Julian, Bergemann, Pugh, Steiner, Alireza, McMullan, Nakamura, Haselwimmer,
  Walker $et$~$al$.}}]{Forsythe2002PRL}
\bibinfo{author}{\bibfnamefont{D.}~\bibnamefont{Forsythe}}
  \bibnamefont{$et$~$al$.}, \bibinfo{journal}{Phys. Rev. Lett.}
  \textbf{\bibinfo{volume}{89}}, \bibinfo{pages}{166402}
  (\bibinfo{year}{2002}).

\bibitem[{\citenamefont{Mackenzie $et$~$al$.}(1998)\citenamefont{Mackenzie,
  Haselwimmer, Tyler, Lonzarich, Mori, Nishizaki, and
  Maeno}}]{Mackenzie1998PRL}
\bibinfo{author}{\bibfnamefont{A.~P.} \bibnamefont{Mackenzie}}
  \bibnamefont{$et$~$al$.}, \bibinfo{journal}{Phys. Rev. Lett.}
  \textbf{\bibinfo{volume}{80}}, \bibinfo{pages}{161} (\bibinfo{year}{1998}).

\bibitem[{\citenamefont{Mao $et$~$al$.}(1999)\citenamefont{Mao, Mori, and
  Maeno}}]{Mao1999PRB}
\bibinfo{author}{\bibfnamefont{Z.~Q.} \bibnamefont{Mao}}
  \bibnamefont{$et$~$al$.}, \bibinfo{journal}{Phys. Rev. B}
  \textbf{\bibinfo{volume}{60}}, \bibinfo{pages}{610} (\bibinfo{year}{1999}).

\bibitem[{\citenamefont{Maeno $et$~$al$.}(1998)\citenamefont{Maeno, Ando, Mori,
  Ohmichi, Ikeda, NishiZaki, and Nakatsuji}}]{Maeno1998PRL}
\bibinfo{author}{\bibfnamefont{Y.}~\bibnamefont{Maeno}}
  \bibnamefont{$et$~$al$.}, \bibinfo{journal}{Phys. Rev. Lett.}
  \textbf{\bibinfo{volume}{81}}, \bibinfo{pages}{3765} (\bibinfo{year}{1998}).

\bibitem[{\citenamefont{Ando $et$~$al$.}(1999)\citenamefont{Ando, Akima, Mori,
  and Maeno}}]{Ando1999JPSJ}
\bibinfo{author}{\bibfnamefont{T.}~\bibnamefont{Ando}}
  \bibnamefont{$et$~$al$.}, \bibinfo{journal}{J. Phys. Soc. Jpn.}
  \textbf{\bibinfo{volume}{68}}, \bibinfo{pages}{1651} (\bibinfo{year}{1999}).

\bibitem[{\citenamefont{Kittaka
  $et$~$al$.}(2009{\natexlab{a}})\citenamefont{Kittaka, Nakamura, Yaguchi,
  Yonezawa, and Maeno}}]{Kittaka2009JPSJ}
\bibinfo{author}{\bibfnamefont{S.}~\bibnamefont{Kittaka}}
  \bibnamefont{$et$~$al$.}, \bibinfo{journal}{J. Phys. Soc. Jpn.}
  \textbf{\bibinfo{volume}{78}}, \bibinfo{pages}{064703}
  (\bibinfo{year}{2009}{\natexlab{a}}).

\bibitem[{\citenamefont{Nobukane $et$~$al$.}(2009)\citenamefont{Nobukane,
  Inagaki, Ichimura, Yamaya, Takayanagi, Kawasaki, Tenya, Amitsuka, Konno,
  Asano $et$~$al$.}}]{Nobukane2009SSC}
\bibinfo{author}{\bibfnamefont{H.}~\bibnamefont{Nobukane}}
  \bibnamefont{$et$~$al$.}, \bibinfo{journal}{Solid State Commun.}
  \textbf{\bibinfo{volume}{149}}, \bibinfo{pages}{1212} (\bibinfo{year}{2009}).

\bibitem[{\citenamefont{Yaguchi $et$~$al$.}(2003)\citenamefont{Yaguchi, Wada,
  Akima, Maeno, and Ishiguro}}]{Yaguchi2003PRB}
\bibinfo{author}{\bibfnamefont{H.}~\bibnamefont{Yaguchi}}
  \bibnamefont{$et$~$al$.}, \bibinfo{journal}{Phys. Rev. B}
  \textbf{\bibinfo{volume}{67}}, \bibinfo{pages}{214519}
  (\bibinfo{year}{2003}).

\bibitem[{\citenamefont{Kawamura $et$~$al$.}(2005)\citenamefont{Kawamura,
  Yaguchi, Kikugawa, Maeno, and Takayanagi}}]{Kawamura2005JPSJ}
\bibinfo{author}{\bibfnamefont{M.}~\bibnamefont{Kawamura}}
  \bibnamefont{$et$~$al$.}, \bibinfo{journal}{J. Phys. Soc. Jpn.}
  \textbf{\bibinfo{volume}{74}}, \bibinfo{pages}{531} (\bibinfo{year}{2005}).

\bibitem[{\citenamefont{Okuda $et$~$al$.}(2002)\citenamefont{Okuda, Suzuki,
  Mao, Maeno, and Fujita}}]{Okuda2002JPSJ}
\bibinfo{author}{\bibfnamefont{N.}~\bibnamefont{Okuda}}
  \bibnamefont{$et$~$al$.}, \bibinfo{journal}{J. Phys. Soc. Jpn.}
  \textbf{\bibinfo{volume}{71}}, \bibinfo{pages}{1134} (\bibinfo{year}{2002}).

\bibitem[{\citenamefont{Nomura and Yamada}(2002)}]{Nomura2002JPSJ-2}
\bibinfo{author}{\bibfnamefont{T.}~\bibnamefont{Nomura}} \bibnamefont{and}
  \bibinfo{author}{\bibfnamefont{K.}~\bibnamefont{Yamada}},
  \bibinfo{journal}{J. Phys. Soc. Jpn.} \textbf{\bibinfo{volume}{71}},
  \bibinfo{pages}{1993} (\bibinfo{year}{2002}).

\bibitem[{\citenamefont{Kittaka
  $et$~$al$.}(2009{\natexlab{b}})\citenamefont{Kittaka, Yaguchi, and
  Maeno}}]{Kittaka2009JPSJ-2}
\bibinfo{author}{\bibfnamefont{S.}~\bibnamefont{Kittaka}}
  \bibnamefont{$et$~$al$.}, \bibinfo{journal}{J. Phys. Soc. Jpn.}
  \textbf{\bibinfo{volume}{78}}, \bibinfo{pages}{103705}
  (\bibinfo{year}{2009}{\natexlab{b}}).

\bibitem[{\citenamefont{Yaguchi $et$~$al$.}(2009)\citenamefont{Yaguchi,
  Kittaka, and Maeno}}]{Yaguchi2009JPCS}
\bibinfo{author}{\bibfnamefont{H.}~\bibnamefont{Yaguchi}}
  \bibnamefont{$et$~$al$.}, \bibinfo{journal}{J. Phys.: Conf. Ser.}
  \textbf{\bibinfo{volume}{150}}, \bibinfo{pages}{052285}
  (\bibinfo{year}{2009}).

\bibitem[{\citenamefont{Mao $et$~$al$.}(2000)\citenamefont{Mao, Maeno, and
  Fukazawa}}]{Mao2000MRB}
\bibinfo{author}{\bibfnamefont{Z.~Q.} \bibnamefont{Mao}}
  \bibnamefont{$et$~$al$.}, \bibinfo{journal}{Mat. Res. Bull.}
  \textbf{\bibinfo{volume}{35}}, \bibinfo{pages}{1813} (\bibinfo{year}{2000}).

\bibitem[{\citenamefont{Smith and Chu}(1967)}]{Smith1967PR}
\bibinfo{author}{\bibfnamefont{T.~F.} \bibnamefont{Smith}} \bibnamefont{and}
  \bibinfo{author}{\bibfnamefont{C.~W.} \bibnamefont{Chu}},
  \bibinfo{journal}{Phys. Rev.} \textbf{\bibinfo{volume}{159}},
  \bibinfo{pages}{353} (\bibinfo{year}{1967}).

\bibitem[{\citenamefont{Sigrist and Monien}(2001)}]{Sigrist2001JPSJ}
\bibinfo{author}{\bibfnamefont{M.}~\bibnamefont{Sigrist}} \bibnamefont{and}
  \bibinfo{author}{\bibfnamefont{H.}~\bibnamefont{Monien}},
  \bibinfo{journal}{J. Phys. Soc. Jpn.} \textbf{\bibinfo{volume}{70}},
  \bibinfo{pages}{2409} (\bibinfo{year}{2001}).

\bibitem[{\citenamefont{Ying $et$~$al$.}(2009)\citenamefont{Ying, Xin, Clouser,
  Hao, Staley, Myers, Allard, Fobes, Liu, Mao $et$~$al$.}}]{Ying2009PRL}
\bibinfo{author}{\bibfnamefont{Y.~A.} \bibnamefont{Ying}}
  \bibnamefont{$et$~$al$.}, \bibinfo{journal}{Phys. Rev. Lett.}
  \textbf{\bibinfo{volume}{103}}, \bibinfo{pages}{247004}
  (\bibinfo{year}{2009}).

\end{thebibliography}

\end{document}